\documentclass[a4paper,fleqn]{cas-dc}

\usepackage[numbers, square, sort&compress]{natbib}
\usepackage{float}
\usepackage{stfloats}

\def\tsc#1{\csdef{#1}{\textsc{\lowercase{#1}}\xspace}}
\tsc{WGM}
\tsc{QE}

\begin{document}
\let\WriteBookmarks\relax
\def\floatpagepagefraction{1}
\def\textpagefraction{.001}

\shorttitle{}    

\shortauthors{C. Timms et al.}  

\title [mode = title]{Nonlinear Photoemission for Bright Beams in X-Band Photoinjectors}  

\author[1]{Casey Timms}[orcid=0009-0002-4814-6298]
\cormark[1]
\ead{catimms@asu.edu}
\credit{Formal Analysis, Investigation, Software, Visualization, Writing - original draft}

\author[1]{Lucas Malin}[orcid=0000-0002-6153-1185]
\credit{Investigation, Visualization, Writing - original draft}

\author[1,2]{Kevin Eckrosh}[orcid=0009-0002-1558-8719]
\credit{Investigation, Software, Visualization}

\author[1,2]{Sean E. Tilton}[orcid=0009-0008-5350-839X]
\credit{Investigation}

\author[1,2]{Lidia C. Santander}[orcid=0009-0007-4180-0863]
\credit{Investigation, Software}

\author[1,2]{Hyung Seo Lee}[orcid=0009-0008-4129-9324]
\credit{Investigation}

\author[1,2]{Rejul Jaswal}
\credit{Investigation}

\author[1]{Alex Gardeck}[orcid=0009-0004-2305-8456]
\credit{Software, Resources}

\author[1,2]{Eric Everett}
\credit{Investigation}

\author[1]{Alan Dupre}[orcid=0009-0008-8227-7367]
\credit{Software}

\author[1]{Mukhtar Hussain}[orcid=0000-0003-3142-9945]
\credit{Investigation}

\author[1,2]{Sami Tantawi}[orcid=0009-0004-0894-6295]
\credit{Supervision}

\author[1,3]{William Graves}[orcid=0000-0002-6634-0548]
\credit{Conceptualization, Supervision, Project Administration}

\author[1]{Henrik Loos}[orcid=0000-0001-5085-0562]
\credit{Software}

\author[1]{Mark Holl}[orcid=0009-0004-4352-1591]
\credit{Resources, Supervision}

\author[1,2]{Arvinder Sandhu}[orcid=0000-0002-7876-855X]
\credit{Supervision, Project Administration}

\author[1,2,3]{Robert Kaindl}[orcid=0000-0003-3639-5625]
\credit{Methodology}

\author[1,2,3]{Samuel Teitelbaum}[orcid=0000-0002-0812-9832]
\credit{Conceptualization, Methodology, Supervision, Writing - review \& editing}

\affiliation[1]{organization={Biodesign Institute, Arizona State University},
            addressline={727 E. Tyler St.}, 
            city={Tempe},
            citysep={}, 
            postcode={85281}, 
            state={Arizona},
            country={United States}}
            
\affiliation[2]{organization={Department of Physics, Arizona State University},
            addressline={Bateman Physics Science F-wing, Rm 470}, 
            city={Tempe},
            citysep={}, 
            postcode={85281}, 
            state={Arizona},
            country={United States}}

\affiliation[3]{organization={Center for Applied Structural Discovery, Arizona State University},
            addressline={97 E. Tyler St.}, 
            city={Tempe},
            citysep={}, 
            postcode={85281}, 
            state={Arizona},
            country={United States}}

\cortext[1]{Corresponding author}


\begin{abstract}
The ongoing development of low-cost and compact x-ray light sources is vital for increasing the number of experiments that can be performed at the atomic scale. One option for a compact and simple photoinjector is to operate in blow-out mode, in which a high-charge, short-bunch length electron beam self-expands by space-charge forces into a uniformly ellipsoidal shape. This mode necessitates a photoinjector whose initial electron pulse has a pulse duration on the order of $100$ fs. A photoinjector utilizing linear photoemission with an ultraviolet laser in this regime presents significant technical challenges, as many optical substrates degrade at these wavelengths. In this paper, we present an alternative design that uses nonlinear photoemission to produce electron beams with high charge and short bunch length in an X-band (9300 MHz) photoinjector. This photoinjector design produces an electron beam with a bunch charge of up to $200$ pC, an RMS bunch length of $291$ fs, a normalized transverse emittance of $1$ to $1.4$ mm$\cdot$mrad, and an energy spread of $10^{-4}$. This results in a beam brightness of $10^{18}$ A/m$^2$, which is bright enough for our beamline's purposes. We also present ongoing efforts are aimed at further increasing the electron beam brightness by using a spatial light modulator to shape the laser profile at the cathode. 
\end{abstract}




\maketitle


\section{Introduction}\label{sec1}
The production of high brightness electron beams \cite{musumeci_advances_2018, moody_longitudinal_2009} with ultrashort bunches is critical to the development of the next generation of free electron lasers \cite{carbone_perspective_2012}, synchrotrons \cite{sedigh_rahimabadi_review_2020}, and ultrafast electron diffraction beamlines \cite{sciaini_femtosecond_2011}. Such facilities have driven paradigm shifts across many fields of the physical and life sciences by allowing experiments to study ultrafast dynamics at the atomic scale \cite{bloembergen_nanosecond_1999, rousse_femtosecond_2001}. In order to allow more researchers and experiments to take advantage of these technologies, it is critical that lower-cost and compact sources are developed. As part of this effort, the Compact X-ray Light Source (CXLS) at Arizona State University aims to create a high-brightness, short pulse duration x-ray beam from a beamline that is only $10$ meters long; much smaller than a typical x-ray free electron laser or synchrotron \cite{graves_compact_2014}.

CXLS generates ultrashort hard x-ray pulses though inverse Compton scattering (ICS), in which the light is emitted from the collision between a relativistic electron bunch collides and a high intensity laser pulse \cite{catravas_femtosecond_2001, hartemann_compton_2007}. The pulse duration of x-rays generated from ICS is approximately equal to the electron beam bunch length \cite{kashiwagi_observation_2000}. To produce an ultrashort electron bunch, CXLS operates in blow-out mode, in which the space charge forces create a self-similar uniform ellipsoid in phase space that can be re-compressed at the interaction point\cite{shamuilov_emittance_2022}. This mode requires that the electron bunch be short at the time of emission from the photocathode. The number of generated x-rays is proportional to the number of electrons in the laser beam focal volume, so a high bunch charge (approximately $200$ pC), focused to a small area, is required \cite{sweers_optimizing_2026}.

To produce an electron bunch, most x-ray sources based on copper linac technology employ linear, or single-photon, photoemission from a copper cathode \cite{alley_design_1999, zhang_linac_2024}. In this regime, the energy from one ultraviolet (UV) photon is absorbed by an electron, allowing it to overcome the work function of the copper cathode and be freed from the material \cite{fowler_analysis_1931}. The pulse duration of these injector lasers is typically greater than $1$ ps, and RF compression techniques are used later in the beamline to shorten the electron bunch \cite{dowell_magnetic_1995, rimjaem_femtosecond_2004}. This long pulse duration is favorable as nonlinear effects and laser-induced damage are less prevalent for laser pulses with lower peak power. Unlike typical beamlines however, CXLS requires that the bunch length at the time of emission be less than $200$ fs, so the photoinjector laser must have an equally short pulse duration.

Initially we attempted to replicate the traditional single photon photoinjectors by illuminating a copper cathode with a $257.5$ nm laser with a pulse duration of $150$ fs, a pulse energy of $48$ $\mu$J, and a repetition rate of $1000$ Hz. However, this design produced numerous unexpected challenges. The optics used to transport the laser to the cathode dropped in transmission over time, often in ways that are challenging to reproduce. We attribute this effect to some combination of laser induced damage and radiation damage (Sec \ref{sec2}). Because our laser has a shorter pulse duration and higher repetition rate than other comparable photoinjectors, we speculate that our optics are more susceptible laser induced damage at UV wavelengths and aging effects occur much faster. For example, optics that would take a year to show damage from a $120$ Hz, $3$ ps laser, would show damage in only a month from a $1000$ Hz, $150$ fs laser. 

Typically, optical substrates have a higher laser induced damage threshold (LIDT) at longer wavelengths \cite{gallais_wavelength_2015}, so we were motivated to design a photoinjector which uses $515$ nm laser light instead of $257.5$ nm by using nonlinear photoemission. The use of nonlinear photoemission for an x-band photoinjector was first demonstrated by \cite{musumeci_multiphoton_2010}. 

\begin{figure}
  \centering
    \includegraphics[width=\linewidth]{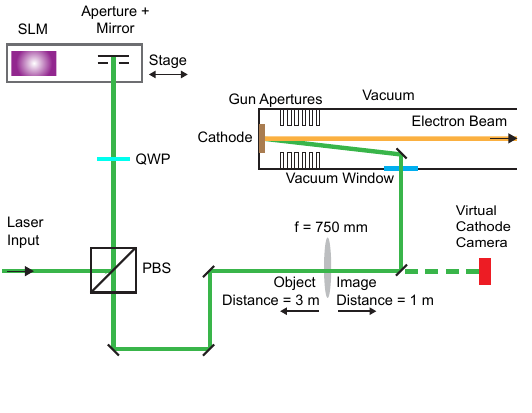}
    \caption{Optical design of the nonlinear photoinjector. The $515$ nm laser light is first reflected by the polarizing beam splitter (PBS), and becomes circularly polarized by the quarter waveplate (QWP) Reflection off of the mirror returns a pulse with unchanged intensity. Reflection off of the SLM returns a pulse whose intensity is related to the phase retardance introduced by the SLM pixels. A stage is used to switch between the mirror and the SLM. A $750$ mm focal length lens images either an aperture placed directly before the mirror or the surface of the SLM onto the cathode. The object distance is $3$ m and the image distance is $1$ m such that the magnification fo the system is $1/3$. A virtual cathode camera monitors the beam profile on the cathode.}\label{optical_design}
\end{figure}

In this work, we present the design (Fig \ref{optical_design}) and experimental validation of a nonlinear photoemission scheme used in our  photoinjector which produces $200$ pC of charge and a 6D beam brightness of $9.81\times 10^{19}$ A/mm$^2$. In addition to eliminating UV-induced optics damage, the switch to visible light allowed us to use a spatial light modulator (SLM) to shape the transverse electron beam profile. Using this device, we can shape the laser pulse to remove nonuniformities from the electron beam and in principle, decrease the root mean square (RMS) emittance. Overall, these results demonstrate that nonlinear photoemission can be used to generate bright electron beams from a simple optical layout that avoids UV-induced optics damage.

\subsection{Expected Charge from Nonlinear Photoemission}\label{sec1.1}
The generalized Fowler-Dubridge theory of photoemission uses Fermi-Dirac statistics to conclude that the total emitted current density $J$ is equal to the sum of partial current densities $J_n$, with 
\begin{equation}J_n = a_n\left[\frac{e}{h\nu}(1-R)I(\textbf{r},t)\right]^nA_0T_e^2f(x_n)\label{Jn}\end{equation}
where $f(x_n)=\int^{\infty}_0\ln(1+\text{exp}(\frac{nh\nu-\phi}{k_BT_e}-y))dy$ is the Fowler function. $a_n$ are phenomenological constants, $h\nu$ is the photon energy, $R$ is the reflectivity of the material, $I(\textbf{r},t)$ is the laser intensity, $A_0$ is the Richardson constant \cite{crowell_richardson_1965}, $T_e$ is the electron temperature, $\phi$ is the work function of the material, and $k_B$ is the Boltzmann constant \cite{fowler_analysis_1931, dubridge_theory_1933, bechtel_two-photon_1977}. Here, $J_n$ represents the contribution from an $n$-photon process. The constants $a_n$ are associated with the probability that an electron will absorb $n$-photons and escape the material, and are related by the ratio: $a_n/a_{n+1}=10^{14}$ A/cm$^2$ \cite{peloi_non-linear_2001}. Additionally, $f(x_n)$ is much greater for $nh\nu>\phi$ than $nh\nu<\phi$, so the dominant current density contribution comes from the lowest $J_n$ term for which \begin{equation}nh\nu>\phi\label{n_condition}\end{equation} Thus, for a fixed laser pulse duration, the charge density $\sigma$ is given by \begin{equation}\sigma \propto I^n\label{sigma_I_relation}\end{equation} where $n$ is the lowest integer that satisfies Eq \ref{n_condition}. So, if we use a copper cathode with work function $\phi=4.31$ eV and a $515$ nm laser with photon energy $h\nu=2.407$ eV, we should expect $\sigma\propto I^2$.

The first electrons emitted from the cathode surface will exert a repulsive force on the later emitted electrons, driving them back into the material \cite{riffe_femtosecond_1993}. Eventually, the field from the extracted charge cancels the accelerating field of the gun and photoemission is suppressed. The charge density at which the accelerating field, $G_a$, is canceled, is given by \cite{pasmans_extreme_2016}: \begin{equation}\sigma_{lim}=\epsilon_0\beta G_a \label{charge_limit} \end{equation} where $\beta$ is a constant related to the field enchantment due to roughness of the cathode surface ($\beta=1$ for a flat cathode). If $I_{lim}$ is the laser intensity for which $\sigma=\sigma_{lim}$, and $I\geq I_{lim}$, then the emitted charge density is proportional to the log of the laser intensity \cite{rosenzweig_initial_1994}. Combining this with Eq \ref{sigma_I_relation} yields\begin{equation}\sigma(I) = \begin{cases}b\cdot I^n & I\leq I_{lim} \\ b\cdot I_{lim}^n+\sigma_{lim}\ln{(I/I_{lim})} & I \geq I_{lim}\end{cases} \label{charge_scaling}\end{equation} where $b$ is a constant that depends on the cathode material and surface structure, laser wavelength, and RF gun field geometry. We use this function to fit the intensity dependence of our photoemission data with $b$, $n$, and $\sigma_{lim}$ as fit parameters. For our system, at low intensities, we expect the charge density to be proportional to the square of the incident laser intensity. At high intensities, we can expect the charge density to be proportional to the log of the laser intensity. 

\begin{figure*}[width = \textwidth]
  \centering
    \includegraphics[width = \textwidth]{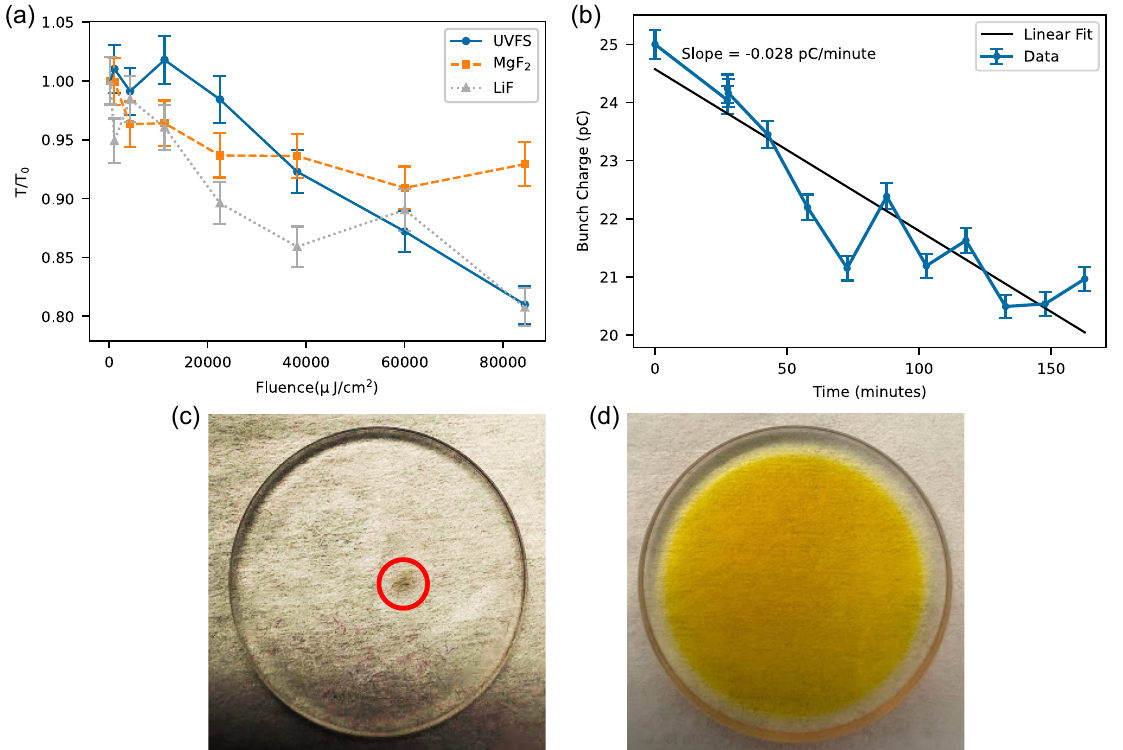}
    \caption{(a) Transmission of optical substrates at $257$ nm after exposure to increasing laser fluence from a $257.5$ nm laser in a test setup. Exposure at high fluences lasted a few seconds. (b) Bunch charged produced by the photoinjector over 150 minutes. (c) Photograph of a MgF$_2$ window after use in the photoinjector. A suspected burn mark is circled in red. (d) Photograph of a LiF window after use in the photoinjector. The yellow color on the outer surface is atypical and believed to be caused by ionizing radiation and high energy electron exposure.}\label{uv_damage}
\end{figure*}

\section{UV Laser Damage with Femtosecond Pulses}\label{sec2}
Table \ref{design_goals} defines the design goals for the CXLS photoinjector. The initial design used linear photoemission from a copper cathode, using a $257.5$ nm central wavelength pulse created by generating the fourth harmonic of a $1030$ nm Yb:KGW laser via two sequential doubling stages (1030 nm to 532 nm, then 532 nm to 257 nm) using BBO crystals. The laser (Light Conversion Pharos) has a pulse duration of $150$ fs, repetition rate of $1000$ Hz, and a pulse energy up to $160$ $\mu$J (Table \ref{uv_laser}).

\begin{table}
\caption{CXLS Photoinjector Design Goals}\label{design_goals}
\begin{tabular*}{\tblwidth}{@{}LL@{}}
\toprule
\textbf{Parameter} & \textbf{Goal} \\ 
\midrule
    Bunch Charge & $200$ pC \\
    Bunch Shape & Ellipsoidal \\
    Maximum Diameter & $2$ mm \\
    Bunch Length at Emission & $<200$ fs \\
    Repetition Rate & $1000$ Hz \\
\bottomrule
\end{tabular*}
\end{table}

\begin{table}
\caption{UV Laser Parameters}\label{uv_laser}
\begin{tabular*}{\tblwidth}{@{}LL@{}}
\toprule
\textbf{Parameter} & \textbf{Value} \\ 
\midrule
    Wavelength & $257.5$ nm \\
    Pulse Duration & $150$ fs \\
    Repetition Rate & $1000$ Hz \\
    Maximum Pulse Energy & $160$ $\mu$J \\
\bottomrule
\end{tabular*}
\end{table}

The QE of the copper cathode in a vacuum pressure of $10^{-8}$ Torr was measured to be approximately $2\times10^{-5}$. This implies a pulse energy of $48.3$ $\mu$J is required to produce $200$ pC of bunch charge. For a $2$ mm diameter beam, this equates to a fluence of approximately $1.50$ mJ/cm$^2$. To ensure that the vacuum window before the cathode could withstand this fluence, we measured the damage threshold of three commonly used optical substrates for UV pulses: Corning 7980 UV fused silica (UVFS), magnesium fluoride (MgF$_2$), and lithium fluoride (LiF) (Fig \ref{uv_damage} (a)). For each uncoated substrate, we measured the initial transmission (T$_0$) of the substrate before exposing it to  high laser intensity at $257.5$ nm for a few seconds, and remeasuring the transmission (T). After each trial, the transmission was remeasured with low UV power to distinguish permanent damage from nonlinear absorption or other reversible nonlinear effects. All three substrates exhibited damage after exposure to $2.2\times10^4$ $\mu$J/cm$^2$; however, MgF$_2$ performed the best, as it retained nearly $95$\% of its transmissivity after exposure to $8.4\times10^4$ $\mu$J/cm$^2$. Therefore, when used in the photoinjector, windows made from MgF$_2$ should not exhibit any significant loss of transmissivity.

While using a MgF$_2$ window in the photoinjector, a pulse energy of $6$ $\mu$J and a laser fluence of $192$ $\mu$J/cm$^2$ was used to to produce a bunch charge of $25$ pC (Fig \ref{uv_damage} (b)). We continued to measure the bunch charge every $15$ minutes for $2.5$ hours, and observed that the bunch charge decreased to $21$ pC. This equates to a pulse energy of $5.1$ $\mu$J, meaning the transmissivity of the window dropped to $85$\% of its initial value over the course of the experiment. Moreover, these data show a charge loss of $0.028$ pC/minute, which is not sustainable for facility that is intended to operate continuously for several days without service. After removing the window, we observed a burn mark (Fig \ref{uv_damage} (c)), which indicates that the loss in transmissivity is due in part to laser induced damage. 

We repeated this experiment with LiF vacuum windows, but observed a more rapid decrease in charge. Rather than exhibiting a burn mark however, the window took on a yellowish color on the vacuum-facing side (Fig \ref{uv_damage} (d)). This is likely due to the window's close proximity to the radiation cone of the electron beam, as ionizing radiation is known to cause strong color center formation in LiF \cite{mclaughlin_electron_1979}. 

We are unable to explain why damage was observed at much lower fluence on the MgF$_2$ window while it was used in the photoinjector. The radiation damage in the LiF window however, drives us to speculate that the ionizing radiation produced by the electron beam decreases the LIDT of the substrate. Alternatively, it may be that any damage occurs only over a long period of time and was therefore not measurable during the instantaneous damage experiments. Regardless of the damage mechanism, we have shown that using a $257.5$ nm laser with a sub-picosecond pulse duration and high repetition rate leaves the photoinjector system uniquely vulnerable to laser induced damage not typically observed at larger photoinjectors operating at low reprate.

From this experience, we derive a few lessons learned for building UV photoinjector laser systems for compact sources. First, although blowout mode is attractive for its simplicity and lack of requirements on longitudinal pulse shaping, special attention must be paid to peak intensities and UV absorption. UV pulses with sub-picosecond pulse durations may have nearly unavoidable laser-induced damage. Second, selection of window substrates is challenging due to a combination of needing to avoid both optical damage and radiation damage. LiF has a high optical damage threshold but is very sensitive to electron radiation, while UVFS is radiation-resistant, but sensitive to intense UV absorption. In addition to nonlinear photoemission, other potential solutions include placing the vacuum entrance window far away from the main accelerator path to reduce radiation damage, and using an in-vacuum optics line with telescopes to reduce the peak intensity on any entrance windows while maintaining a constant beam size on the cathode surface.

\section{Nonlinear Photoemission}\label{sec3}
We chose to investigate nonlinear photoemission using $515$ nm light, generated via second harmonic generation (SHG) of our $1030$ nm Yb:KGW laser. This is approach is advantageous as the LIDT of optical substrates tends to increase with wavelength \cite{gallais_wavelength_2015}. The new laser parameters are given in Table \ref{green_laser}. We created an ellipsoidal beam at the cathode by imaging a pinhole aperture onto the cathode surface (Fig \ref{optical_design}). The focal lengths of the imaging lens was $750$ mm, the object distance was $3$ m, and the image distance was $1$ m such that electron beam diameter was $1/3$ of the diameter of the aperture. In order to avoid self focusing caused by focusing a high-intensity beam through air, a vacuum was placed between the lenses. The lenses and window substrates were MgF$_2$, due to the low nonlinear index of this material at $515$ nm \cite{patwardhan_nonlinear_2021}. A virtual cathode camera (VCC) was placed after a beamsplitter such that it was an equal optical distance as the cathode from the focusing lens to monitor the laser profile in the imaging plane.

\begin{table}
\caption{Green Laser Parameters}\label{green_laser}
\begin{tabular*}{\tblwidth}{@{}LL@{}}
\toprule
\textbf{Parameter} & \textbf{Value} \\ 
\midrule
    Wavelength & $515$ nm \\
    Pulse Duration & $150$ fs \\
    Repetition Rate & $1000$ Hz \\
    Maximum Pulse Energy & $540$ $\mu$J \\
\bottomrule
\end{tabular*}
\end{table}

\begin{figure}
  \centering
    \includegraphics[width = \linewidth]{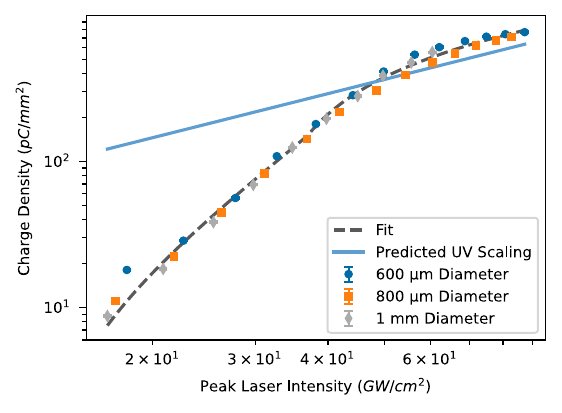}
    \caption{Emitted charge density plotted as a function of laser intensity. At low intensity, data is fitted to a third-order power law, and at high intensity data is fitted to a natural log. The expected charge density from linear photoemission is also plotted, and shown to be less than the charge density produced by nonlinear photoemission at high intensity. Note that error bars are not visible due to the logarithmic scaling.}\label{charge_data}
\end{figure}

\begin{figure}
  \centering
    \includegraphics[width = \linewidth]{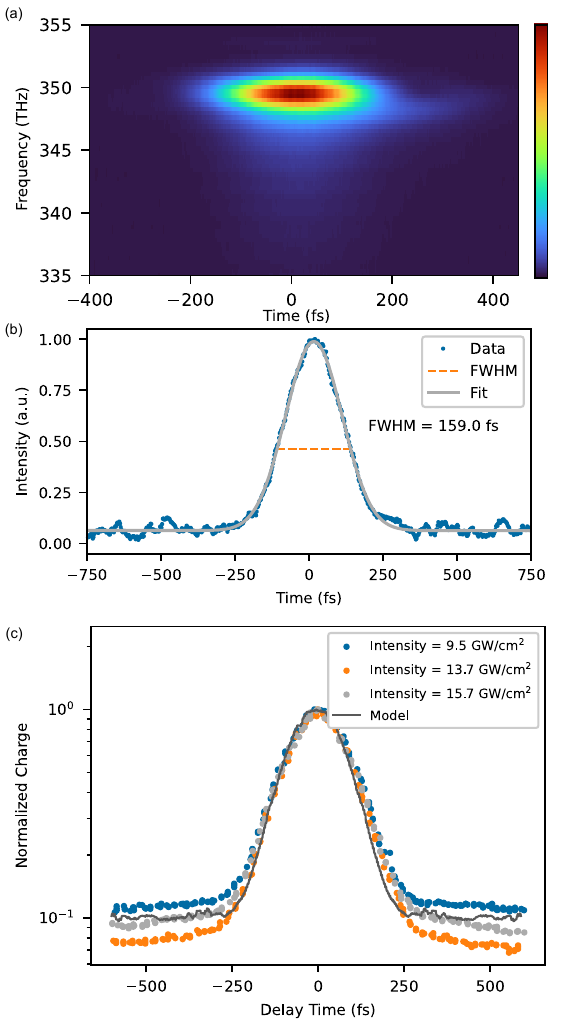}
    \caption{(a) Frequency Resolved Optical Gating (FROG) trace of the laser pulse taken using a second harmonic generation FROG. (b) Optical autocorrelation of the laser pulse derived from the FROG trace. The full width at half maximum (FWHM) is shown to be $159$ fs. (c) Charge autocorrelations at different laser intensities taken by combining two identical laser pulses at the cathode and delaying one relative to the other. A model of the charge autocorrelation is calculated as the autocorrelation of the cube of two identical pulses with a FWHM of $159$ fs. Adherence of autocorrelations to the model suggests that thermal effects do not play a role in the emission.}\label{autocorrelation}
\end{figure}

We measured the bunch charge as a function of laser intensity for three different beam diameters and fit the data to Eq \ref{charge_scaling} (Fig \ref{charge_data}). Note that these data were collected using a different imaging system, which utilized two lenses to produce an image of the aperture at the cathode with $2/3$ magnification. Although we expect that $\sigma\propto I^2$ at low intensities, (sec \ref{sec1.1}), we observed that $\sigma \propto I^{3}$. This increased power-law scaling has been previously observed by \cite{fujimoto_femtosecond_1984} in tungsten cathodes and \cite{bormann_tip-enhanced_2010} in metal nanotips. This suggests that the Fowler-Dubridge approach for determining the emitted current by using Fermi-Dirac statistics to calculate the number of electrons with energy above the cathode work function is incomplete. \cite{yalunin_strong-field_2011} also considers not only the probability that an electron absorbs enough energy to overcome the work function, but also considers the probability that a free electron reflects off of the cathode surface and back into the material. Their results suggest that the increased power-law scaling is due to this chance of the back reflection. Additionally, it may be that there are a higher number of electrons available at lower energies, causing the three-photon process to become more favorable.

While  the charge density followed a power law at low laser intensities, at laser intensities above $35$ GW/cm$^2$, the charge density was proportional to the natural log of the laser intensity. This is in agreement with the scaling shown in Eq \ref{charge_scaling}. At high intensities, the charge density threshold is measured to be $\sigma_{lim} = 744$ pC/mm$^2$, which agrees with Eq \ref{charge_limit} for our gun field of $G_a=65.4$ MV/m. Comparison of the charge density produced by nonlinear photoemission to the charge density that would be produced by linear photoemission from a cathode with $\text{QE}=2\times10^{-5}$ shows that nonlinear photoemission is more efficient for laser intensities greater than $45$ GW/cm$^2$.

Previous studies of nonlinear photoemission have shown that at high laser intensities, the electron gas can have a significantly higher temperature than the lattice \cite{ferrini_non-linear_2009, fujimoto_femtosecond_1984}. The increased electron temperature causes enhanced emission long after the laser pulse arrives at the cathode, effectively increasing the electron bunch length. To show that thermal effects are not playing a significant role in the emission process, we first measured the laser pulse duration using a SHG frequency resolved optical gating (FROG) (Fig \ref{autocorrelation} (a)). We extracted the optical autocorrelation from the FROG trace and found that the pulse has a full width at half maximum (FWHM) of $159$ fs (Fig \ref{autocorrelation} (b)). Using this FWHM, we simulated the expected charge autocorrelation for third-order photoemission (Fig \ref{autocorrelation} (c)). We measured the charge autocorrelation by splitting the photoinjector laser into two identical pulses and delaying one pulse relative to the other. The pulses were recombined on the cathode, and we observed that when the pulses were temporally overlapped, the charge produced was an order of magnitude greater than the charge produced by two non-overlapped pulses (Fig \ref{autocorrelation} (c)). We measured autocorrelations over a range of laser intensities from $9.5$ GW/cm$^2$ to $15.7$ GW/cm$^2$, and found that the pulse shape did not change as the laser intensity increased. Moreover, all of the measured autocorrelations adhered closely to the simulated charge autocorrelation. This indicates that thermal effects do not play a role in the emission process, and are thus not affecting the bunch length.

\begin{figure}
  \centering
    \includegraphics[width = \linewidth]{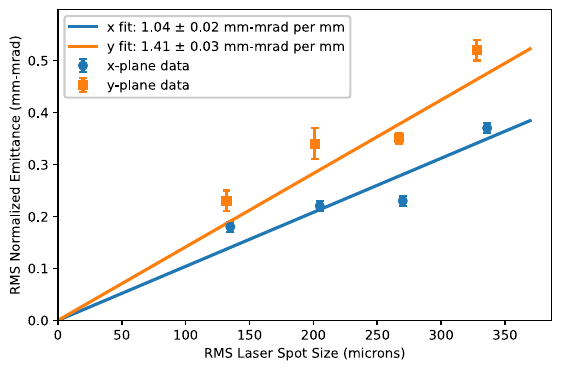}
    \caption{Measurement of the thermal emittance of the electron bunch along the x and y dimensions.}\label{emittance}
\end{figure}

The 6D brightness of the electron beam is a conserved quantity as the electron beam passes through steering and focusing optics, and contributes greatly to the overall performance of the beamline \cite{davut_balance_2025}. This quantity is given as
\begin{equation}
B_{6D}=\frac{Q}{\varepsilon_x \varepsilon_y \sigma_z \Delta E/E}
\label{6D_brightness}
\end{equation}
where $Q$ is the bunch charge, $\varepsilon_x$, $\varepsilon_y$, are the RMS normalized emittances along the x and y directions respectively, $\sigma_z$ is the RMS bunch length, and $\Delta E/E$ is the energy spread. From a $1$ mm diameter beam, we measured a peak bunch charge of $200$ pC. Figure \ref{emittance} shows the emittance growth with the laser spot size, so for a $1$ mm diameter beam, $\varepsilon_x=1.04$ mm-mrad and $\varepsilon_y=1.41$ mm-mrad. For this beam size, the energy spread was measured to be $\Delta E/E=10^{-4}$.

\begin{figure}
  \centering
    \includegraphics[width = \linewidth]{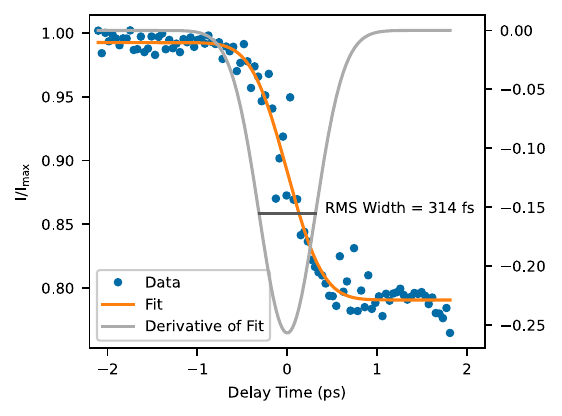}
    \caption{Transmitted laser intensity through scintillator as a function of delay time. Data is fitted to an error function and the derivative of the fit is plotted. The RMS bunch length is taken to be the standard deviation of this gaussian fit.}\label{bunch_length}
\end{figure}

To measure the bunch length, we used a technique similar to the one described by \cite{eckrosh_characterization_2025}. Under this approach, the electron beam is used to pump a YAG:Ce crystal, causing a decrease in the transmission of the IR light through the material. By propagating a $64$ fs (RMS) $1030$ nm laser pulse with the electron beam, we can measure the transmissivity of the YAG as a function of delay relative to the electron bunch (Fig \ref{bunch_length}). Because the transmission effect lasts for several nanoseconds, we observe an error function-like response. The derivative of this error function is taken to be the convolution of the temporal shape of the electron bunch, the temporal shape of the IR laser, the response time of the YAG, and the walkoff between the electron bunch and laser. This walkoff is caused by the fact that the group velocity of IR will be slower in the YAG than the velocity of the electron beam, which is nearly the vacuum speed of light. Therefore, the RMS width of the derivative ($\sigma_{d}$) is given by
\begin{equation}
\sigma_d^2 = \sigma_e^2+\sigma_{\text{IR}}^2+\sigma_{\text{YAG}}^2+\sigma_{\text{walkoff}}^2
\label{convolution}
\end{equation}
where $\sigma_e$ is the RMS bunch length of the electron beam, $\sigma_{\text{IR}}$ is the RMS width of the IR pulse, $\sigma_{\text{YAG}}$ is the response time of the YAG, and $\sigma_{\text{walkoff}}$ is the walkoff caused by the difference in travel time between the electron bunch and IR pulse. Deconvolving these effects yields
\begin{equation}
\sigma_e=\sqrt{\sigma_d^2-\sigma_{\text{IR}}^2-\sigma_{\text{YAG}}^2-\sigma_{\text{walkoff}}^2}
\label{deconvolution}
\end{equation}

We measured the RMS width of the derivative to be $314$ fs, and the RMS width of the IR pulse was $64$ fs. The response time of the YAG is taken to be approximately $70$ fs \cite{sato_simple_2019}. YAG has a refractive index of $1.82$ at $1030$ nm, so for a $100$ $\mu$m thick crystal, the difference in travel time between the electron bunch and IR pulse is $272$ fs \cite{zelmon_refractive-index_1998}. So, the walkoff is approximately $68$ fs. Thus, from Eq \ref{deconvolution}, we can conclude that the bunch length of the electron beam is $291$ fs. From Eq \ref{6D_brightness}, the 6D beam brightness is approximately $5\times10^{18}$ A/m$^2$, which is sufficiently high to achieve a bright x-ray beam using ICS. Overall, this result demonstrates that for beamlines operating in blowout mode, nonlinear photoemission is highly favorable for the production of bright electron beams.

\begin{figure*}[width = \textwidth]
  \centering
    \includegraphics[width = \textwidth]{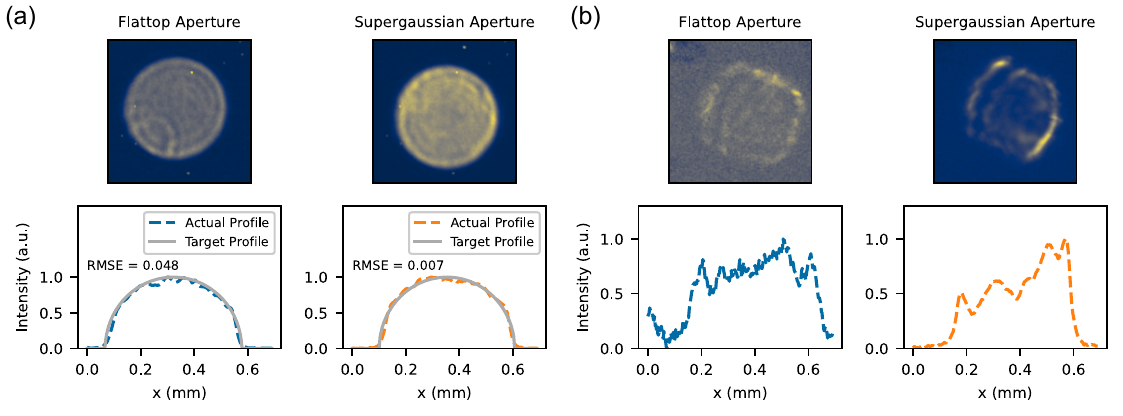}
    \caption{(a) Comparison of laser profiles created using a flattop aperture (left) to a super gaussian aperture (right). Linecuts of both profiles show that the beam created with the super gaussian aperture adheres more closely to the ideal ellipsoidal shape than the beam created using the flattop aperture. (b) Comparison of the electron beam profiles created using a flattop aperture (left) to the super gaussian aperture (right). Both profiles display significant nonuniformities not present in the laser profiles, which indicates that the cathode surface introduces structure into the electron beam profile}\label{slm_profiles}
\end{figure*}

\section{Polarization Masking}\label{sec4}
It is desirable to further increase the beam brightness by making the electron bunch adhere more closely to the ideal ellipsoidal profile, and thereby decrease its transverse emittance \cite{kim_emittance_1997, luiten_how_2004}. This ideal intensity is given as:
\begin{equation}
I_{\text{ideal}}(r)=I_{\text{max}}\sqrt{1-(r/R)^2}
\label{ellipse}
\end{equation}
where $I_{\text{max}}$ is the peak intensity, $r$ is the radial coordinate, and $R$ is the beam radius. Using nonlinear photoemission opens up the opportunity to use a spatial light modulator (SLM) to adaptively shape the beam's transverse profile and achieve this ideal profile, as there are many commercially available SLMs compatible with $515$ nm light. This adaptive beam shaping can be achieved by implementing an SLM alongside a polarization masking system. Using this system, the portion of light reflected off of an SLM pixel has an intensity given by \cite{ye_construction_1995, davis_two-dimensional_2000}:
\begin{equation}
I=I_0\sin^2(\phi/2)
\label{slm_intensity}
\end{equation}
where $I_0$ is the incident intensity of the laser light and $\phi$ is the phase retardance introduced by the pixel. Thus, by replacing the aperture in Fig \ref{optical_design} with an SLM, and imaging the surface of the SLM onto the cathode, we can create arbitrary laser profiles that are transferred to the produced electron bunch. Such a design has been demonstrated by \cite{maxson_adaptive_2015}, and will allow us to correct for nonuniformities introduced by the optics, cathode surface, or misalignment through the system.

After implementation of the SLM, the first step was to correct for diffraction effects introduced by hard-edged apertures. Although these effects should disappear when the aperture is imaged onto the cathode, the limited numerical apertures of the lenses, windows, and gun as well as misalignment of the lenses can cause diffraction rings to persist. Because the photoemission is nonlinear, the system is much more sensitive to edge diffraction than UV photoinjectors. To correct for the diffraction edges, we used the SLM to change the effective beam shape from a sharp flattop profile to a tenth-order supergaussian. We measured the resultant laser profiles at VCC and compared these measured profiles to the ideal profile (Eq \ref{ellipse}).

For a measured intensity profile $I_{\text{actual}}$, the root mean squared error was calculated to be
\begin{equation}
\text{RMSE}=\sqrt{\frac{1}{N}\sum_i^N(I_{\text{actual}}(r_i)-I_{\text{ideal}}(r_i))^2}
\label{RMSE}
\end{equation}
where $i$ is the pixel number and $N$ is the number of pixels where either $I_{\text{actual}}$ or $I_{\text{ideal}}$ is nonzero. We found that while the profile created using a flattop aperture had an RMSE of 0.048, the profile created using the supergaussian aperture had an RMSE of just 0.007 (Fig \ref{slm_profiles} (a)). This indicates that even without adaptive shaping, using an SLM to generate a supergaussian aperture can be used to increase the uniformity of the laser profile.

Further optimizing the beam profile requires directly imaging the electron beam transverse profile at the cathode. We measured the electron beam profiles at the cathode by using a solenoid magnet to image the cathode surface onto a downstream scintillator screen  located $135$ mm from the cathode (Fig \ref{slm_profiles} (b)). Both of the profiles exhibited strong nonuniformities not present in the laser profiles, indicating the cathode surface itself is introducing structure to the beam. Ongoing work is aimed at using adaptive beam shaping algorithms to remove this structure. For instance, \cite{maxson_adaptive_2015} takes an iterative approach in which the phase written to each SLM pixel is modified from the previous phase iteration based on the difference between the targeted electron beam profile and the measured profile. Using such an approach, we hope to produce a uniformly ellipsoidal electron bunch with low emittance.

\section{Conclusions}\label{sec5}
We have demonstrated that photoinjectors which aim to create ultrashort electron bunches from a copper cathode can benefit from using nonlinear photoemission. This is because the high-intensity UV light required for linear photoemission tends to damage and degrade optical substrates over long periods of time. These substrates are more robust for longer wavelengths however, so, in cases where the laser pulses have high intensity, nonlinear photoinjectors offer a relatively simple alternative. We have shown that at high laser intensities, the charge density produced by a nonlinear photoinjector is comparable to that of a linear photoinjector, and we have demonstrated a 6D beam brightness of approximately $5\times10^{18}$ A/m$^2$. Additionally, nonlinear photoinjectors allow for the implementation of SLMs, which can be used to improve the uniformity of the laser profile. Future work is aimed at correcting nonuniformities in the electron beam profile to create a bunch with very low emittance. Overall, these results demonstrate that nonlinear photoemission can be used to create bright electron beams from a simple, low-cost design.




\printcredits

\section*{Acknowledgements}
This material was based upon work supported by the National Science Foundation under Grant Nos. 2153503 and 1935994.

\bibliographystyle{unsrtnat}

\bibliography{main_bib}

@article{mclaughlin_electron_1979,
	title = {Electron and gamma-ray dosimetry using radiation-induced color centers in {LiF}},
	volume = {14},
	issn = {0146-5724},
	url = {https://www.sciencedirect.com/science/article/pii/0146572479900839},
	doi = {10.1016/0146-5724(79)90083-9},
	number = {3},
	urldate = {2026-05-26},
	journal = {Radiation Physics and Chemistry (1977)},
	author = {McLaughlin, W. L. and Lucas, A. C. and Kapsar, B. M. and Miller, A.},
	month = jan,
	year = {1979},
	pages = {467--480},
}

@article{gallais_wavelength_2015,
	title = {Wavelength dependence of femtosecond laser-induced damage threshold of optical materials},
	volume = {117},
	issn = {0021-8979},
	url = {https://doi.org/10.1063/1.4922353},
	doi = {10.1063/1.4922353},
	number = {22},
	urldate = {2026-05-26},
	journal = {Journal of Applied Physics},
	author = {Gallais, L. and Douti, D.-B. and Commandr{\'e}, M. and Batavi{\v c}i{\=u}t{\.e}, G. and Pupka, E. and {\v S}{\v c}iuka, M. and Smalakys, L. and Sirutkaitis, V. and Melninkaitis, A.},
	month = jun,
	year = {2015},
	pages = {223103},
}

@article{fujimoto_femtosecond_1984,
	title = {Femtosecond {Laser} {Interaction} with {Metallic} {Tungsten} and {Nonequilibrium} {Electron} and {Lattice} {Temperatures}},
	volume = {53},
	url = {https://link.aps.org/doi/10.1103/PhysRevLett.53.1837},
	doi = {10.1103/PhysRevLett.53.1837},
	number = {19},
	urldate = {2026-05-26},
	journal = {Physical Review Letters},
	publisher = {American Physical Society},
	author = {Fujimoto, J. G. and Liu, J. M. and Ippen, E. P. and Bloembergen, N.},
	month = nov,
	year = {1984},
	pages = {1837--1840},
}

@article{bormann_tip-enhanced_2010,
	title = {Tip-{Enhanced} {Strong}-{Field} {Photoemission}},
	volume = {105},
	url = {https://link.aps.org/doi/10.1103/PhysRevLett.105.147601},
	doi = {10.1103/PhysRevLett.105.147601},
	number = {14},
	urldate = {2026-05-26},
	journal = {Physical Review Letters},
	publisher = {American Physical Society},
	author = {Bormann, R. and Gulde, M. and Weismann, A. and Yalunin, S. V. and Ropers, C.},
	month = sep,
	year = {2010},
	pages = {147601},
}

@article{yalunin_strong-field_2011,
	title = {Strong-field photoemission from surfaces: {Theoretical} approaches},
	volume = {84},
	shorttitle = {Strong-field photoemission from surfaces},
	url = {https://link.aps.org/doi/10.1103/PhysRevB.84.195426},
	doi = {10.1103/PhysRevB.84.195426},
	number = {19},
	urldate = {2026-05-26},
	journal = {Physical Review B},
	publisher = {American Physical Society},
	author = {Yalunin, Sergey V. and Gulde, Max and Ropers, Claus},
	month = nov,
	year = {2011},
	pages = {195426},
}

@article{maxson_adaptive_2015,
	title = {Adaptive electron beam shaping using a photoemission gun and spatial light modulator},
	volume = {18},
	url = {https://link.aps.org/doi/10.1103/PhysRevSTAB.18.023401},
	doi = {10.1103/PhysRevSTAB.18.023401},
	number = {2},
	urldate = {2026-05-26},
	journal = {Physical Review Special Topics - Accelerators and Beams},
	publisher = {American Physical Society},
	author = {Maxson, Jared and Lee, Hyeri and Bartnik, Adam C. and Kiefer, Jacob and Bazarov, Ivan},
	month = feb,
	year = {2015},
	pages = {023401},
}

@article{patwardhan_nonlinear_2021,
	title = {Nonlinear refractive index of solids in mid-infrared},
	volume = {46},
	copyright = {{\textcopyright} 2021 Optical Society of America},
	issn = {1539-4794},
	url = {https://opg.optica.org/ol/abstract.cfm?uri=ol-46-8-1824},
	doi = {10.1364/OL.421469},
	language = {EN},
	number = {8},
	urldate = {2026-05-26},
	journal = {Optics Letters},
	publisher = {Optica Publishing Group},
	author = {Patwardhan, Gauri N. and Ginsberg, Jared S. and Chen, Cecilia Y. and Jadidi, M. Mehdi and Gaeta, Alexander L.},
	month = apr,
	year = {2021},
	pages = {1824--1827},
}

@article{crowell_richardson_1965,
	title = {The {Richardson} constant for thermionic emission in {Schottky} barrier diodes},
	volume = {8},
	issn = {0038-1101},
	url = {https://www.sciencedirect.com/science/article/pii/0038110165901164},
	doi = {10.1016/0038-1101(65)90116-4},
	number = {4},
	urldate = {2026-05-26},
	journal = {Solid-State Electronics},
	author = {Crowell, C. R.},
	month = apr,
	year = {1965},
	pages = {395--399},
}

@article{bechtel_two-photon_1977,
	title = {Two-photon photoemission from metals induced by picosecond laser pulses},
	volume = {15},
	copyright = {http://link.aps.org/licenses/aps-default-license},
	issn = {0556-2805},
	url = {https://link.aps.org/doi/10.1103/PhysRevB.15.4557},
	doi = {10.1103/PhysRevB.15.4557},
	language = {en},
	number = {10},
	urldate = {2026-05-27},
	journal = {Physical Review B},
	author = {Bechtel, J. H. and Lee Smith, W. and Bloembergen, N.},
	month = may,
	year = {1977},
	pages = {4557--4563},
}

@article{musumeci_multiphoton_2010,
	title = {Multiphoton {Photoemission} from a {Copper} {Cathode} {Illuminated} by {Ultrashort} {Laser} {Pulses} in an rf {Photoinjector}},
	volume = {104},
	url = {https://link.aps.org/doi/10.1103/PhysRevLett.104.084801},
	doi = {10.1103/PhysRevLett.104.084801},
	number = {8},
	urldate = {2026-05-27},
	journal = {Physical Review Letters},
	publisher = {American Physical Society},
	author = {Musumeci, P. and Cultrera, L. and Ferrario, M. and Filippetto, D. and Gatti, G. and Gutierrez, M. S. and Moody, J. T. and Moore, N. and Rosenzweig, J. B. and Scoby, C. M. and Travish, G. and Vicario, C.},
	month = feb,
	year = {2010},
	pages = {084801},
}

@article{peloi_non-linear_2001,
	title = {Non-linear photoemission from polycrystalline molybdenum irradiated by 790 nm{\textendash}150 ~ fs laser pulses},
	volume = {118},
	issn = {0038-1098},
	url = {https://www.sciencedirect.com/science/article/pii/S0038109801001302},
	doi = {10.1016/S0038-1098(01)00130-2},
	number = {7},
	urldate = {2026-05-27},
	journal = {Solid State Communications},
	author = {Peloi, Marco and Ferrini, Gabriele and Banfi, GianPiero and Secondi, Gianluca and Parmigiani, Fulvio},
	month = may,
	year = {2001},
	pages = {339--344},
}

@article{fowler_analysis_1931,
	title = {The {Analysis} of {Photoelectric} {Sensitivity} {Curves} for {Clean} {Metals} at {Various} {Temperatures}},
	volume = {38},
	url = {https://link.aps.org/doi/10.1103/PhysRev.38.45},
	doi = {10.1103/PhysRev.38.45},
	number = {1},
	urldate = {2026-05-27},
	journal = {Physical Review},
	publisher = {American Physical Society},
	author = {Fowler, R. H.},
	month = jul,
	year = {1931},
	pages = {45--56},
}

@article{dubridge_theory_1933,
	title = {Theory of the {Energy} {Distribution} of {Photoelectrons}},
	volume = {43},
	url = {https://link.aps.org/doi/10.1103/PhysRev.43.727},
	doi = {10.1103/PhysRev.43.727},
	number = {9},
	urldate = {2026-05-27},
	journal = {Physical Review},
	publisher = {American Physical Society},
	author = {DuBridge, Lee A.},
	month = may,
	year = {1933},
	pages = {727--741},
}

@article{riffe_femtosecond_1993,
	title = {Femtosecond thermionic emission from metals in the space-charge-limited regime},
	volume = {10},
	copyright = {{\textcopyright} 1993 Optical Society of America},
	issn = {1520-8540},
	url = {https://opg.optica.org/josab/abstract.cfm?uri=josab-10-8-1424},
	doi = {10.1364/JOSAB.10.001424},
	language = {EN},
	number = {8},
	urldate = {2026-05-27},
	journal = {JOSA B},
	publisher = {Optica Publishing Group},
	author = {Riffe, D. M. and Wang, X. Y. and Downer, M. C. and Fisher, D. L. and Tajima, T. and Erskine, J. L. and More, R. M.},
	month = aug,
	year = {1993},
	pages = {1424--1435},
}

@article{pasmans_extreme_2016,
	title = {Extreme regimes of femtosecond photoemission from a copper cathode in a dc electron gun},
	volume = {19},
	url = {https://link.aps.org/doi/10.1103/PhysRevAccelBeams.19.103403},
	doi = {10.1103/PhysRevAccelBeams.19.103403},
	number = {10},
	urldate = {2026-05-27},
	journal = {Physical Review Accelerators and Beams},
	publisher = {American Physical Society},
	author = {Pasmans, P. L. E. M. and van Vugt, D. C. and van Lieshout, J. P. and Brussaard, G. J. H. and Luiten, O. J.},
	month = oct,
	year = {2016},
	pages = {103403},
}

@article{rosenzweig_initial_1994,
	title = {Initial measurements of the {UCLA} rf photoinjector},
	volume = {341},
	issn = {0168-9002},
	url = {https://www.sciencedirect.com/science/article/pii/0168900294903875},
	doi = {10.1016/0168-9002(94)90387-5},
	number = {1},
	urldate = {2026-05-27},
	journal = {Nuclear Instruments and Methods in Physics Research Section A: Accelerators, Spectrometers, Detectors and Associated Equipment},
	author = {Rosenzweig, J. and Barov, N. and Hartman, S. and Hogan, M. and Park, S. and Pellegrini, C. and Travish, G. and Zhang, R. and Davis, P. and Hairapetian, G. and Joshi, C.},
	month = mar,
	year = {1994},
	pages = {379--385},
}

@article{kim_emittance_1997,
	title = {Emittance in particle and radiation beam techniques},
	volume = {398},
	issn = {0094-243X},
	url = {https://doi.org/10.1063/1.53053},
	doi = {10.1063/1.53053},
	number = {1},
	urldate = {2026-05-27},
	journal = {AIP Conference Proceedings},
	author = {Kim, Kwang-Je},
	month = mar,
	year = {1997},
	pages = {243--253},
}

@article{luiten_how_2004,
	title = {How to {Realize} {Uniform} {Three}-{Dimensional} {Ellipsoidal} {Electron} {Bunches}},
	volume = {93},
	url = {https://link.aps.org/doi/10.1103/PhysRevLett.93.094802},
	doi = {10.1103/PhysRevLett.93.094802},
	number = {9},
	urldate = {2026-05-28},
	journal = {Physical Review Letters},
	publisher = {American Physical Society},
	author = {Luiten, O. J. and van der Geer, S. B. and de Loos, M. J. and Kiewiet, F. B. and van der Wiel, M. J.},
	month = aug,
	year = {2004},
	pages = {094802},
}

@article{davis_two-dimensional_2000,
	title = {Two-dimensional polarization encoding with a phase-only liquid-crystal spatial light modulator},
	volume = {39},
	copyright = {{\textcopyright} 2000 Optical Society of America},
	issn = {2155-3165},
	url = {https://opg.optica.org/ao/abstract.cfm?uri=ao-39-10-1549},
	doi = {10.1364/AO.39.001549},
	language = {EN},
	number = {10},
	urldate = {2026-05-28},
	journal = {Applied Optics},
	publisher = {Optica Publishing Group},
	author = {Davis, Jeffrey A. and McNamara, Dylan E. and Cottrell, Don M. and Sonehara, Tomio},
	month = apr,
	year = {2000},
	pages = {1549--1554},
}

@article{ye_construction_1995,
	title = {Construction of an optical rotator using quarter-wave plates and an optical retarder},
	volume = {34},
	issn = {0091-3286},
	url = {https://www.spiedigitallibrary.org/journals/optical-engineering/volume-34/issue-10/0000/Construction-of-an-optical-rotator-using-quarter-wave-plates-and/10.1117/12.210734},
	doi = {10.1117/12.210734},
	language = {en},
	number = {10},
	urldate = {2026-05-28},
	journal = {Optical Engineering},
	publisher = {SPIE},
	author = {Ye, Chun},
	month = oct,
	year = {1995},
	pages = {3031--3035},
}

@article{carbone_perspective_2012,
	title = {A perspective on novel sources of ultrashort electron and {X}-ray pulses},
	volume = {392},
	issn = {0301-0104},
	url = {https://www.sciencedirect.com/science/article/pii/S0301010411004319},
	doi = {10.1016/j.chemphys.2011.10.010},
	number = {1},
	urldate = {2026-05-28},
	journal = {Chemical Physics},
	author = {Carbone, F. and Musumeci, P. and Luiten, O. J. and Hebert, C.},
	month = jan,
	year = {2012},
	pages = {1--9},
}

@article{sedigh_rahimabadi_review_2020,
	title = {Review on applications of synchrotron-based {X}-ray techniques in materials characterization},
	volume = {49},
	copyright = {{\textcopyright} 2020 John Wiley \& Sons Ltd},
	issn = {1097-4539},
	url = {https://onlinelibrary.wiley.com/doi/abs/10.1002/xrs.3141},
	doi = {10.1002/xrs.3141},
	language = {en},
	number = {3},
	urldate = {2026-05-28},
	journal = {X-Ray Spectrometry},
	author = {Sedigh Rahimabadi, Pooria and Khodaei, Mehdi and Koswattage, Kaveenga R.},
	year = {2020},
	pages = {348--373},
}

@article{sciaini_femtosecond_2011,
	title = {Femtosecond electron diffraction: heralding the era of atomically resolved dynamics},
	volume = {74},
	issn = {0034-4885},
	shorttitle = {Femtosecond electron diffraction},
	url = {https://doi.org/10.1088/0034-4885/74/9/096101},
	doi = {10.1088/0034-4885/74/9/096101},
	language = {en},
	number = {9},
	urldate = {2026-05-28},
	journal = {Reports on Progress in Physics},
	author = {Sciaini, Germ{\'a}n and Miller, R J Dwayne},
	month = aug,
	year = {2011},
	pages = {096101},
}

@article{bloembergen_nanosecond_1999,
	title = {From nanosecond to femtosecond science},
	volume = {71},
	url = {https://link.aps.org/doi/10.1103/RevModPhys.71.S283},
	doi = {10.1103/RevModPhys.71.S283},
	number = {2},
	urldate = {2026-05-28},
	journal = {Reviews of Modern Physics},
	publisher = {American Physical Society},
	author = {Bloembergen, N.},
	month = mar,
	year = {1999},
	pages = {S283--S287},
}

@article{rousse_femtosecond_2001,
	title = {Femtosecond x-ray crystallography},
	volume = {73},
	url = {https://link.aps.org/doi/10.1103/RevModPhys.73.17},
	doi = {10.1103/RevModPhys.73.17},
	number = {1},
	urldate = {2026-05-28},
	journal = {Reviews of Modern Physics},
	publisher = {American Physical Society},
	author = {Rousse, Antoine and Rischel, Christian and Gauthier, Jean-Claude},
	month = jan,
	year = {2001},
	pages = {17--31},
}

@article{graves_compact_2014,
	title = {Compact x-ray source based on burst-mode inverse {Compton} scattering at 100 {kHz}},
	volume = {17},
	url = {https://link.aps.org/doi/10.1103/PhysRevSTAB.17.120701},
	doi = {10.1103/PhysRevSTAB.17.120701},
	number = {12},
	urldate = {2026-05-28},
	journal = {Physical Review Special Topics - Accelerators and Beams},
	publisher = {American Physical Society},
	author = {Graves, W. S. and Bessuille, J. and Brown, P. and Carbajo, S. and Dolgashev, V. and Hong, K.-H. and Ihloff, E. and Khaykovich, B. and Lin, H. and Murari, K. and Nanni, E. A. and Resta, G. and Tantawi, S. and Zapata, L. E. and K{\"a}rtner, F. X. and Moncton, D. E.},
	month = dec,
	year = {2014},
	pages = {120701},
}

@article{hartemann_compton_2007,
	title = {Compton scattering x-ray sources driven by laser wakefield acceleration},
	volume = {10},
	url = {https://link.aps.org/doi/10.1103/PhysRevSTAB.10.011301},
	doi = {10.1103/PhysRevSTAB.10.011301},
	number = {1},
	urldate = {2026-05-28},
	journal = {Physical Review Special Topics - Accelerators and Beams},
	publisher = {American Physical Society},
	author = {Hartemann, F. V. and Gibson, D. J. and Brown, W. J. and Rousse, A. and Phuoc, K. Ta and Mallka, V. and Faure, J. and Pukhov, A.},
	month = jan,
	year = {2007},
	pages = {011301},
}

@article{catravas_femtosecond_2001,
	title = {Femtosecond x-rays from {Thomson} scattering using laser wakefield accelerators},
	volume = {12},
	issn = {0957-0233},
	url = {https://doi.org/10.1088/0957-0233/12/11/310},
	doi = {10.1088/0957-0233/12/11/310},
	language = {en},
	number = {11},
	urldate = {2026-05-28},
	journal = {Measurement Science and Technology},
	author = {Catravas, P. and Esarey, E. and Leemans, W. P.},
	month = oct,
	year = {2001},
	pages = {1828},
}

@article{rimjaem_femtosecond_2004,
	title = {Femtosecond electron bunches from an {RF}-gun},
	volume = {533},
	issn = {0168-9002},
	url = {https://www.sciencedirect.com/science/article/pii/S0168900204014391},
	doi = {10.1016/j.nima.2004.05.135},
	number = {3},
	urldate = {2026-05-28},
	journal = {Nuclear Instruments and Methods in Physics Research Section A: Accelerators, Spectrometers, Detectors and Associated Equipment},
	author = {Rimjaem, Sakhorn and Farias, Ruy and Thongbai, Chitrlada and Vilaithong, Thiraphat and Wiedemann, Helmut},
	month = nov,
	year = {2004},
	pages = {258--269},
}

@inproceedings{dowell_magnetic_1995,
	title = {Magnetic pulse compression using a third harmonic {RF} linearizer},
	volume = {2},
	url = {https://ieeexplore.ieee.org/document/505106/},
	doi = {10.1109/PAC.1995.505106},
	urldate = {2026-05-28},
	booktitle = {Proceedings {Particle} {Accelerator} {Conference}},
	author = {Dowell, D.H. and Hayward, T.D. and Vetter, A.M.},
	month = may,
	year = {1995},
	pages = {992--994 vol.2},
}

@article{zhang_linac_2024,
	title = {The {Linac} {Coherent} {Light} {Source} {II} photoinjector laser infrastructure},
	volume = {12},
	issn = {2095-4719, 2052-3289},
	url = {https://www.cambridge.org/core/journals/high-power-laser-science-and-engineering/article/linac-coherent-light-source-ii-photoinjector-laser-infrastructure/3598F0E668472B10241B04F3536186A6},
	doi = {10.1017/hpl.2024.33},
	language = {en},
	urldate = {2026-05-28},
	journal = {High Power Laser Science and Engineering},
	author = {Zhang, Hao and Gilevich, Sasha and Miahnahri, Alan and Alverson, Shawn Christopher and Brachmann, Axel and Duris, Joseph and Franz, Paris and Fry, Alan and Hirschman, Jack and Larsen, Kirk and Lemons, Randy and Li, Siqi and Lu, Brittany and Marinelli, Agostino and Martinez, Mikael and May, Justin and Milshtein, Erel and Murari, Krishna and Neveu, Nicole and Robinson, Joseph and Schmerge, John and Sun, Linshan and Vecchione, Theodore and Xu, Chengcheng and Zhou, Feng and Carbajo, Sergio},
	month = jan,
	year = {2024},
	pages = {e51},
}

@article{alley_design_1999,
	title = {The design for the {LCLS} {RF} photoinjector},
	volume = {429},
	issn = {0168-9002},
	url = {https://www.sciencedirect.com/science/article/pii/S0168900299000728},
	doi = {10.1016/S0168-9002(99)00072-8},
	number = {1},
	urldate = {2026-05-28},
	journal = {Nuclear Instruments and Methods in Physics Research Section A: Accelerators, Spectrometers, Detectors and Associated Equipment},
	author = {Alley, R. and Bharadwaj, V. and Clendenin, J. and Emma, P. and Fisher, A. and Frisch, J. and Kotseroglou, T. and Miller, R. H. and Palmer, D. T. and Schmerge, J. and Sheppard, J. C. and Woodley, M. and Yeremian, A. D. and Rosenzweig, J. and Meyerhofer, D. D. and Serafini, L.},
	month = jun,
	year = {1999},
	pages = {324--331},
}

@misc{sweers_optimizing_2026,
	title = {Optimizing the interaction geometry of inverse {Compton} scattering x-ray sources},
	url = {http://arxiv.org/abs/2512.20356},
	doi = {10.48550/arXiv.2512.20356},
	urldate = {2026-05-28},
	publisher = {arXiv},
	author = {Sweers, C. W. and Luiten, O. J.},
	month = feb,
	year = {2026},
	note = {arXiv:2512.20356 [physics.acc-ph]},
}

@article{shamuilov_emittance_2022,
	title = {Emittance self-compensation in blow-out mode},
	volume = {24},
	issn = {1367-2630},
	url = {https://doi.org/10.1088/1367-2630/aca5ab},
	doi = {10.1088/1367-2630/aca5ab},
	language = {en},
	number = {12},
	urldate = {2026-05-28},
	journal = {New Journal of Physics},
	publisher = {IOP Publishing},
	author = {Shamuilov, Georgii and Opanasenko, Anatoliy and Pepitone, K{\'e}vin and Tibai, Zolt{\'a}n and Goryashko, Vitaliy},
	month = dec,
	year = {2022},
	pages = {123008},
}

@article{kashiwagi_observation_2000,
	series = {Proceedings of the {Int}. {Symp}. on {New} {Visions} in {Laser}-{Beam}},
	title = {Observation of high-intensity {X}-rays in inverse {Compton} scattering experiment},
	volume = {455},
	issn = {0168-9002},
	url = {https://www.sciencedirect.com/science/article/pii/S0168900200006896},
	doi = {10.1016/S0168-9002(00)00689-6},
	number = {1},
	urldate = {2026-05-28},
	journal = {Nuclear Instruments and Methods in Physics Research Section A: Accelerators, Spectrometers, Detectors and Associated Equipment},
	author = {Kashiwagi, S and Washio, M and Kobuki, T and Kuroda, R and Ben-Zvi, I and Pogorelsky, I and Kusche, K and Skaritka, J and Yakimenko, V and Wang, X. J and Hirose, T and Dobashi, K and Muto, T and Urakawa, J and Omori, T and Okugi, T and Tsunemi, A and Liu, Y and He, P and Cline, D and Segalov, Z},
	month = nov,
	year = {2000},
	pages = {36--40},
}

@article{moody_longitudinal_2009,
	title = {Longitudinal phase space characterization of the blow-out regime of rf photoinjector operation},
	volume = {12},
	url = {https://link.aps.org/doi/10.1103/PhysRevSTAB.12.070704},
	doi = {10.1103/PhysRevSTAB.12.070704},
	number = {7},
	urldate = {2026-06-10},
	journal = {Physical Review Special Topics - Accelerators and Beams},
	publisher = {American Physical Society},
	author = {Moody, J. T. and Musumeci, P. and Gutierrez, M. S. and Rosenzweig, J. B. and Scoby, C. M.},
	month = jul,
	year = {2009},
	pages = {070704},
}

@article{musumeci_advances_2018,
	series = {Advances in {Instrumentation} and {Experimental} {Methods} ({Special} {Issue} in {Honour} of {Kai} {Siegbahn})},
	title = {Advances in bright electron sources},
	volume = {907},
	issn = {0168-9002},
	url = {https://www.sciencedirect.com/science/article/pii/S0168900218303541},
	doi = {10.1016/j.nima.2018.03.019},
	urldate = {2026-06-10},
	journal = {Nuclear Instruments and Methods in Physics Research Section A: Accelerators, Spectrometers, Detectors and Associated Equipment},
	author = {Musumeci, P. and Giner Navarro, J. and Rosenzweig, J. B. and Cultrera, L. and Bazarov, I. and Maxson, J. and Karkare, S. and Padmore, H.},
	month = nov,
	year = {2018},
	pages = {209--220},
}

@article{ferrini_non-linear_2009,
	series = {Special issue in honour of {Prof}. {Kai} {Siegbahn}},
	title = {Non-linear electron photoemission from metals with ultrashort pulses},
	volume = {601},
	issn = {0168-9002},
	url = {https://www.sciencedirect.com/science/article/pii/S0168900208020184},
	doi = {10.1016/j.nima.2008.12.107},
	number = {1},
	urldate = {2026-06-10},
	journal = {Nuclear Instruments and Methods in Physics Research Section A: Accelerators, Spectrometers, Detectors and Associated Equipment},
	author = {Ferrini, Gabriele and Banfi, Francesco and Giannetti, Claudio and Parmigiani, Fulvio},
	month = mar,
	year = {2009},
	pages = {123--131},
}

@article{davut_balance_2025,
	title = {Balance of bunch compression and emittance preservation for high-brightness x-ray free electron laser injectors},
	volume = {28},
	url = {https://link.aps.org/doi/10.1103/rjns-vqzt},
	doi = {10.1103/rjns-vqzt},
	number = {9},
	urldate = {2026-06-12},
	journal = {Physical Review Accelerators and Beams},
	publisher = {American Physical Society},
	author = {Davut, C. and Apsimon, O. and Hounsell, B. R. and Militsyn, B. L. and Cowie, L. S. and Yaman, F. and Brynes, A. D. and Williams, P. H.},
	month = sep,
	year = {2025},
	pages = {091602},
}

@article{eckrosh_characterization_2025,
	title = {Characterization of spatiotemporal overlap of femtosecond lasers and electron beam with {Ce}:{YAG} screens},
	volume = {96},
	issn = {0034-6748},
	shorttitle = {Characterization of spatiotemporal overlap of femtosecond lasers and electron beam with {Ce}},
	url = {https://doi.org/10.1063/5.0276436},
	doi = {10.1063/5.0276436},
	number = {12},
	urldate = {2026-06-12},
	journal = {Review of Scientific Instruments},
	author = {Eckrosh, Kevin and Tilton, Sean and Malin, Lucas and Brown, Taryn and Dupre, Alan and Semaan, Antonella and Gardeck, Alex and Babic, Gregory and Lee, Hyung Seo and Loos, Henrik and Hussain, Mukhtar and Sandhu, Arvinder and Graves, William S. and Holl, Mark R. and Teitelbaum, Samuel W.},
	month = dec,
	year = {2025},
	pages = {123016},
}

@article{zelmon_refractive-index_1998,
	title = {Refractive-index measurements of undoped yttrium aluminum garnet from 0.4 to 5.0 $\mu$m},
	volume = {37},
	copyright = {{\textcopyright} 1998 Optical Society of America},
	issn = {2155-3165},
	url = {https://opg.optica.org/ao/abstract.cfm?uri=ao-37-21-4933},
	doi = {10.1364/AO.37.004933},
	language = {EN},
	number = {21},
	urldate = {2026-07-07},
	journal = {Applied Optics},
	publisher = {Optica Publishing Group},
	author = {Zelmon, David E. and Small, David L. and Page, Ralph},
	month = jul,
	year = {1998},
	pages = {4933--4935},
}

@article{sato_simple_2019,
	title = {A simple instrument to find spatiotemporal overlap of optical/{X}-ray light at free-electron lasers},
	volume = {26},
	issn = {1600-5775},
	url = {https://journals.iucr.org/s/issues/2019/03/00/ig5070/},
	doi = {10.1107/S1600577519002248},
	language = {en},
	number = {3},
	urldate = {2026-10-12},
	journal = {Journal of Synchrotron Radiation},
	publisher = {International Union of Crystallography},
	author = {Sato, T. and Glownia, J. M. and Ware, M. R. and Chollet, M. and Nelson, S. and Zhu, D.},
	month = may,
	year = {2019},
	pages = {647--652},
}



\end{document}